\documentclass[aps,twocolumn,showpacs,preprintnumbers,amsmath,amssymb,superscriptaddress,floatfix,nofootinbib]{revtex4}

\usepackage{graphicx}
\usepackage{epsfig}
\usepackage{epstopdf}
\usepackage{hyperref}
\usepackage{amsmath}
\usepackage{amsfonts}
\usepackage{amssymb}
\usepackage{dsfont}

\usepackage{graphicx,color,dcolumn,booktabs}
\usepackage{txfonts}
\usepackage{amssymb}
\usepackage{indentfirst}
\usepackage{subfig}
\usepackage{float}
\usepackage{tabularx} % for tabularx

\allowdisplaybreaks
\begin{document}

\title{States of $\rho B^* \bar{B}^*$ with $J=3$ within the Fixed Center
Approximation to Faddeev equations.}

\author{M.~Bayar}
\affiliation{Department of Physics, Kocaeli University, 41380, Izmit, 
Turkey and \\
Departamento de
F\'{\i}sica Te\'orica and IFIC, Centro Mixto Universidad de
Valencia-CSIC Institutos de Investigaci\'on de Paterna, Aptdo.
22085, 46071 Valencia, Spain
}

\author{P.~ Fernandez-Soler}
\affiliation{Departamento de
F\'{\i}sica Te\'orica and IFIC, Centro Mixto Universidad de
Valencia-CSIC Institutos de Investigaci\'on de Paterna, Aptdo.
22085, 46071 Valencia, Spain}

\author{Zhi-Feng Sun}
\affiliation{Departamento de
F\'{\i}sica Te\'orica and IFIC, Centro Mixto Universidad de
Valencia-CSIC Institutos de Investigaci\'on de Paterna, Aptdo.
22085, 46071 Valencia, Spain}

%\author{J. Nieves}
%\affiliation{IFIC, Centro Mixto Universidad de
%Valencia-CSIC Institutos de Investigaci\'on de Paterna, Aptdo.
%22085, 46071 Valencia, Spain}

\author{E.~Oset}
\affiliation{Departamento de
F\'{\i}sica Te\'orica and IFIC, Centro Mixto Universidad de
Valencia-CSIC Institutos de Investigaci\'on de Paterna, Aptdo.
22085, 46071 Valencia, Spain}

\date{\today}

\begin{abstract}
In this work we study the $\rho B^*\bar{B}^*$ three-body system solving the Faddeev equations in the fixed center approximation. We assume the $B^*\bar{B}^*$ system forming a cluster, and in terms of the two-body $\rho B^*$ unitarized scattering amplitudes in the local Hidden Gauge approach we find a new $I(J^{PC})=1(3^{--})$ state. The mass of the new state corresponds to a two particle invariant mass of the $\rho B^\ast$ system close to the resonant energy of the $B^\ast_2(5747)$, indicating that the role of this $J=2$ resonance is important in the dynamical generation of the new state. 
\end{abstract}

\maketitle
\section{Introduction}
The three body problem is a classical one and the formal solution is given in terms of the Faddeev equations \cite{Faddeev:1960su}. Although formally these equations are very easy and intuitive, the technical implementation is rather involved and usually some approximations are always done \cite{Alt:1967fx}. Advances in this field using chiral Lagrangians have also been done \cite{Epelbaum:2000mx}. A different approach, using as input unitarized amplitudes obtained from chiral Lagrangians was done in \cite{MartinezTorres:2008gy,Roca:2010tf,MartinezTorres:2009xb,MartinezTorres:2011vh} for states of three mesons, in \cite{MartinezTorres:2007sr,Khemchandani:2008rk,Sun:2011fr} for two mesons and a baryon and in \cite{Bayar:2011qj,Bayar:2012dd,Ikeda:2007nz} for one meson and two baryons. A different approach, using separable potentials fitted to two body data, is also done in \cite{Gal:2011yp} for two mesons and a baryon are studied. In \cite{Barnea:2012qa}, the $\bar{K}NN$, $\bar{K}NNN$ and $\bar{K}\bar{K}NN$ systems with a chiral $\bar{K}N$ interaction and a phenomenological $NN$ potential. In the present work we deal with the interaction of the three-particle system composed of a $\rho$,
 a $B^\ast$ and a $\bar{B}^\ast$ meson. 

The three-body approach has helped to bring more insight in the understanding of different observed resonances, as in the case of the $\pi (1300)$ \cite{Agashe:2014kda}, that was discussed in Ref. \cite{MartinezTorres:2011vh} as a molecular resonance of the coupled system $\pi K\bar{K}$ and $\pi\pi\eta$. In the work of Ref. \cite{Xie:2011uw} the baryonic resonance $\Delta_{\frac{5}{2}^{+}}(2000)$ \cite{Agashe:2014kda} was proposed as a $\pi-(\Delta\rho)$ system. The mesonic states $f_2(1270)$, $\rho_3(1690)$, $f_4(2050)$, $\rho_5(2350)$ and $f_6(2510)$ \cite{Agashe:2014kda} were obtained in Ref. \cite{Roca:2010tf} as multi-$\rho$ states, interpreting all these resonances as molecular states made of two, three and up to six $\rho$ mesons respectively. This strong interaction with many $\rho$ mesons with alligned spins was also investigated in Ref. \cite{YamagataSekihara:2010qk}, and it was found that the $K^\ast(892)$-multi-$\rho$ interaction can dynamically generate the observed $K^\ast_2(1430)$, $K^\ast_3(1780)$, $K^\ast_4(2045)$ and $K^\ast_5(2380)$ states \cite{Agashe:2014kda}.

In this work, as in 
\cite{MartinezTorres:2008gy,MartinezTorres:2007sr,Khemchandani:2008rk,MartinezTorres:2011vh,Bayar:2011qj,Roca:2010tf,MartinezTorres:2009xb,Sun:2011fr,Bayar:2012dd,Ikeda:2007nz,YamagataSekihara:2010qk,Xie:2011uw}, the 
three-body interaction is solved using as input the two-body interaction of the composing hadrons 
given by the chiral unitary approach. This two-body interaction in the present case, is obtained from the lowest order 
Lagrangian which describes the interaction of pseudoscalar plus vector mesons \cite{Bando:1984ej,Bando:1987br,Meissner:1987ge}, called the local Hidden Gauge approach, after the 
unitarization of the tree level amplitudes \cite{Molina:2008jw,Geng:2008gx}. The extension to the charm sector of the local Hidden Gauge approach \cite{Molina:2009eb} and its success in the description of known $D$ states, plus the prediction of the $D(2600)$, which was found later in the experiment of Ref. \cite{delAmoSanchez:2010vq}, motivated the aplication of this formalism to the study of the $D^\ast$-multi-$\rho$ interaction \cite{Xiao:2012dw}, the three-body interaction of the $\rho D^\ast \bar{D}^\ast$ system \cite{Bayar:2015oea} and of the $\rho D \bar{D}$ system \cite{Durkaya:2015wra}. We follow here the analogous aplication of the two-body interaction of the $\rho B^\ast$ \cite{Soler:2015hna} system to solve the three-body interaction of the $\rho B^\ast \bar{B}^\ast$ system.

One of the findings of \cite{Molina:2008jw,Molina:2009ct,Ozpineci:2013qza} is the existence of bound states in spin $J=2$ for the $\rho \rho$, $D^\ast \bar{D}^\ast$ and $B^\ast\bar{B}^\ast$ systems respectively, as a consequence of the strong vector-vector interaction in this sector. The fact that these $J=2$ states are very bound in all the cases, since the binding energy found is $ \gtrsim 100$ MeV, justifies the fixed center approximation to the Faddeev equations (FCA). This approximation is often used \cite{Chand:1962ec,Toker:1981zh,Barrett:1999cw,Deloff:1999gc} and it was also used in the previous works of Refs. \cite{MartinezTorres:2008gy,MartinezTorres:2007sr,Khemchandani:2008rk,Bayar:2011qj,Roca:2010tf,Sun:2011fr,Bayar:2012dd,Xie:2011uw,YamagataSekihara:2010qk,Xiao:2012dw,Bayar:2015oea}. It is justified since two of the particles form a bound cluster that is not much affected by the scattering of the third particle \cite{Bayar:2012dd,MartinezTorres:2010ax}. In the language of quantum field theory, the FCA was identified with the limit of neglecting the three-momentum of the two heavier particles in the propagators, and then it is only the third, the lighter particle, the one who finally interacts with the cluster \cite{Kamalov:2000iy}.

In the following sections we will explain in detail the formalism employed to solve the interaction, and we will show the results that we obtain with this model for the case in which the pair of $B^*\bar{B}^\ast$ mesons are composing a cluster of spin 2 and isospin 0. We will evaluate the sensitivity of this model to the free parameters and some of the different assumptions, and finally, present the conclusions and scope of this work.
\section{Formalism}
\subsection{The Fixed Center Approximation to Faddeev equations \label{Sec:FCA-F-E}}
\begin{figure*}
	\centering
	\includegraphics[scale=0.7]{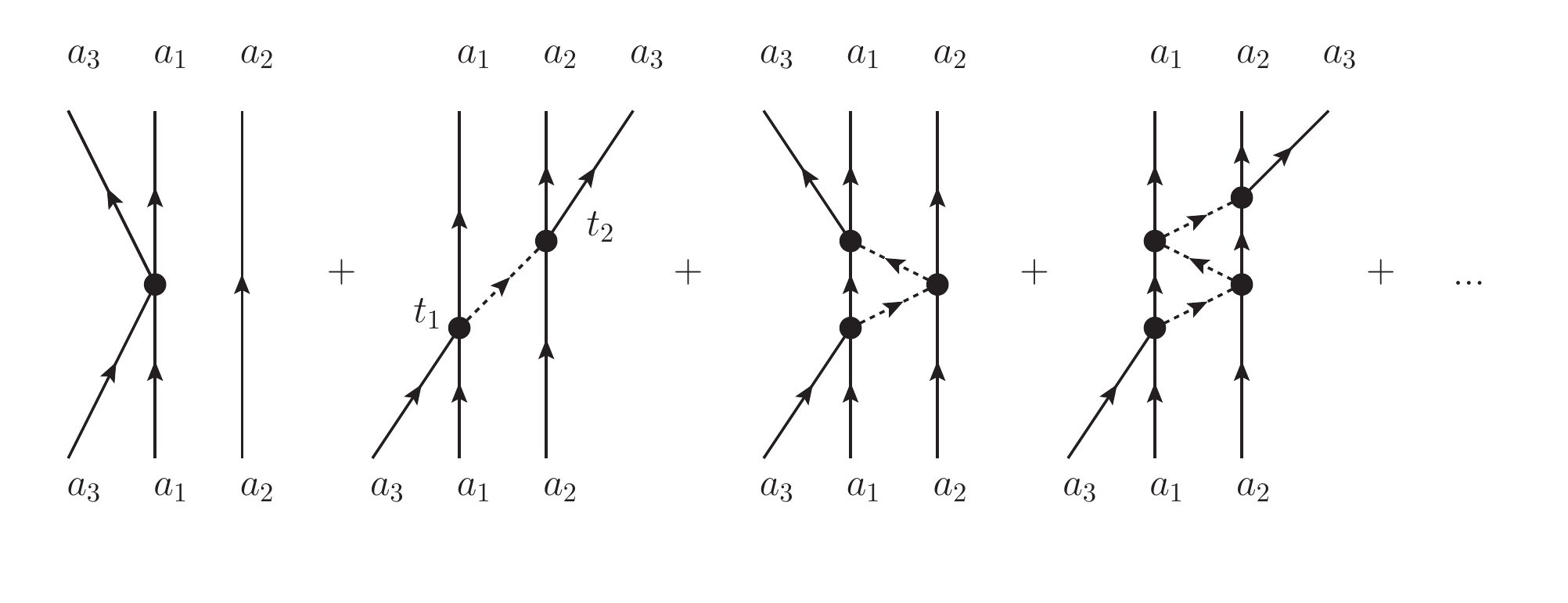}
	\caption{Diagramatic representation of the Faddeev equation resumation
	for $T_1$ in the fixed center approximation.}
	\label{Fig:Scheme-for-T1-2}
\end{figure*}
In the FCA the $B^*$ and $\bar{B}^*$ mesons will be regarded forming a cluster for which we know its wave function. Then we can construct the diagramatic series describing the $\rho$ meson interacting with the cluster, Fig. \ref{Fig:Scheme-for-T1-2}, which can be accounted for in terms of two partition functions, $T_1$, $T_2$, which sum all the multiple scattering terms of particle 3 beginning with particle 1 or 2 respectively. The partition functions satisfy the following coupled equations,
\begin{align}
	T_1=t_1+t_1G_0t_2+t_1 G_0 t_2 G_0 t_1+t_1 G_0 t_2 G_0 t_1 G_0 t_2+\dots  \label{Eq:T_1-FFC-dots},\\
	T_2=t_2+t_2 G_0 t_1+t_2 G_0 t_1 G_0 t_2+ t_2 G_0 t_1 G_0 t_2 G_0 t_1+\dots \label{Eq:T_2-FFC-dots},
\end{align}
taking $t_1 G_0$ and $t_2 G_0$ as a common factor in Eqs. (\ref{Eq:T_1-FFC-dots}) and (\ref{Eq:T_2-FFC-dots}) respectively, we arrive to the coupled Faddeev equations in the fixed center approximation for the particle $a_1$ and $a_2$:
\begin{align}
	T_1=t_1+t_1G_0 T_2 \label{Eq:T_1-FFC},\\
	T_2=t_2+t_2G_0 T_1 \label{Eq:T_2-FFC}.
\end{align}
Now $G_0$ is the propagator for the $\rho$ meson folded with the wave function of the cluster and we will focus on it later. Solving Eqs. \eqref{Eq:T_1-FFC} and \eqref{Eq:T_2-FFC} we will be able to construct the total three-body $T$ matrix,
\begin{align}
	T=T_1+T_2.
	\label{Eq:T-matrix}
\end{align}

In the isospin basis the $\rho$ is a triplet $(-\rho^+,\rho^0,\rho^-)$ and
the $B^*$ and $\bar{B}^*$ are doublets $(B^{*+},B^{*0})$, $(\bar{B}^{*0},-B^{*-})$. 
We will calculate the $\rho(B^*\bar{B}^*)$ interaction in spin $J=3$. 
The $B^*\bar{B}^*$ and $B_s^*\bar{B}_s^*$ systems have been studied previously
in the formalism of the local Hidden Gauge approach and Heavy Quark Spin
Symmetry \cite{Ozpineci:2013qza}. One of the results of this previous work is the absence of bound states of the $B^*\bar{B}^*$ system in isospin $1$, due to the weakness of the interaction in this sector. 
The bound states are found in isospin 0 and we are interested in those of
spin $J_{B^*\bar{B}^*}=2$. We will take the value of the
$B^*\bar{B}^*$ cluster mass for $I(J^{PC})=0\,(2^{++})$ as $M_{\rm c}=10616$ MeV, obtained whithout considering the
coupling to the $B_s^*\bar{B}_s^*$ channel. The consequence of including bottomed strange mesons as a coupled channel is that the mass of the $J=2$ state is obtained as $M_{\rm c}=10613$ MeV \cite{Ozpineci:2013qza}. So the influence of this latter channel can be neglected. These results legitimate the use of this state as a $B^*\bar{B}^*$ bound state.

We will project the scattering amplitudes of the three-body system into proper states of total isospin $1$ in terms of two particle isospin states, with the restriction of considering the $B^*\bar{B}^*$ system as an isospin zero state.
\begin{align}
  \left| \rho B^* \bar{B}^*,I=1,I_3=1\right\rangle =& \left|I=1,I_3=1 \right\rangle_\rho \otimes \left|I=0, I_3=0\right\rangle_{B^*\bar{B}^*} \notag\\
  =&  \left|I=1,I_3=1 \right\rangle_{\rho} \otimes \notag\\
  &\left\lbrace 
    \frac{1}{\sqrt{2}}
    \left( 
    \left|I_3=\frac{1}{2}\right\rangle_{B^*}
    \right.\right. \left.\left.
    \left|I_3=-\frac{1}{2}\right\rangle_{\bar{B}^*} \right.\right.\notag\\
    &\left.\left.-\left|I_3=-\frac{1}{2}\right\rangle_{B^*}
    \left|I_3=\frac{1}{2}\right\rangle_{\bar{B}^*}
    \right)
    \right\rbrace 
    \label{Eq:I-one-three-particle-state}.
\end{align}
When we make the projection 
\begin{align}
  \left\langle \rho B^* \bar{B}^*,I=1\right| T_j\left| \rho B^* \bar{B}^*,I=1\right\rangle,~~ j=1,2,
  \label{Eq:T-matrix-isospin-projection}
\end{align}  
it must be taken into account that 
$t_1$ relates particles $a_1$ and $a_3$, i.e., it describes the $\rho {B}^*$ system. With $t_2$ we are labelling the $T$ matrix for $\rho \bar{B}^*$ interaction. So we need to consider the $\rho B^*$ isospin states:
\begin{align}
    \left|I_{3(\rho)}=1,I_{3({B^*})}=\frac{1}{2}\right\rangle =&\left|I=\frac{3}{2},I_3=\frac{3}{2}\right\rangle_{\rho B^*} \notag\\
    \left|I_{3(\rho)}=1,I_{3({B^*})}=-\frac{1}{2}\right\rangle =&\frac{1}{\sqrt{3}}
    \left|I=\frac{3}{2},I_3=\frac{1}{2}\right\rangle_{\rho B^*}+\notag\\
    & \frac{2}{\sqrt{3}}
    \left|I=\frac{1}{2},I_3=\frac{1}{2}\right\rangle_{\rho B^*}
   \label{Eq:I-rho-Bx-states}.
\end{align}
Combining Eqs. (\ref{Eq:I-one-three-particle-state}), (\ref{Eq:I-rho-Bx-states}) into (\ref{Eq:T-matrix-isospin-projection}), we obtain the following result,
\begin{align}
  t_j(I=1) = \frac{2}{3}\hat{t}^{\,\,I=3/2}_j+\frac{1}{3}\hat{t}^{\,\,I=1/2}_j~~j=1,2.
  \label{Eq:t_j-I-1}
\end{align}
In Eq. (\ref{Eq:t_j-I-1}) $\hat{t}^{\,\,I}_1$ and $\hat{t}^{\,\,I}_2$ are the $T$ matrices of the $\rho {B}^*$ and $\rho \bar{B}^*$ systems respectively, for isospin $I$ of the two particle system. The $\rho B^*$ interaction has already been studied in a previous work \cite{Soler:2015hna}. The $\rho \bar{B}^*$ interaction is the same as that of $\rho B^*$, which could be obtained by charge conjugation.

We are going to calculate the three-body interaction in terms of the two-body $\rho B^*$ results assuming $a_1$ and $a_2$ as a cluster. In order to do it, let us take the first diagram of Fig. \ref{Fig:Scheme-for-T1-2} and write the $S$-matrix element, at first order, for the interaction of $a_3$ with $a_1$ and also of $a_3$ with $a_2$. We follow the convention of \cite{Mandl:1985bg},
\begin{align}
  S_1^{(1)} & =\frac{-{\rm{i}}t_1 }{\sqrt{2\omega(a_3)}\sqrt{2\omega'(a_3)}\sqrt{2\omega(a_1)}\sqrt{2\omega'(a_1)}}\frac{(2\pi)^4\delta^4(P)}{\mathcal{V}^2} \label{Eq:S-Matrix-t1-3-particles}, \\
S_2^{(1)} & =\frac{-{\rm{i}}t_2 }{\sqrt{2\omega(a_3)}\sqrt{2\omega'(a_3)}\sqrt{2\omega(a_2)}\sqrt{2\omega'(a_2)}}\frac{(2\pi)^4\delta^4(P)}{\mathcal{V}^2} \label{Eq:S-Matrix-t2-3-particles} .
\end{align}
The $\delta^4(P)$ ensures the four-momentum conservation, $\omega(a_i)$ and $\omega'(a_i)$ stand for the initial and final energies, respectively, of the $a_i$ particle, and $t_i$ is the amplitude associated to the process, while $\mathcal{V}$ stands for the volume where the states are normalized to 1. 

The $S$-matrix of the $\rho$ meson scattering with the $B^*\bar{B}^*$ cluster of mass $M_{\rm c}$ is expressed as
\begin{align}
 S_c^{(1)} & =\frac{-{\rm{i}}T }{\sqrt{2\omega(a_3)}\sqrt{2\omega'(a_3)}\sqrt{2\omega(\rm{c})}\sqrt{2\omega'(\rm{c})}}\frac{(2\pi)^4\delta^4(P)}{\mathcal{V}^2} \label{Eq:S-Matrix-cluster} .
\end{align}

The factors $(2\omega_i)^{1/2}$ in Eqs. \eqref{Eq:S-Matrix-t1-3-particles}, \eqref{Eq:S-Matrix-t2-3-particles} and \eqref{Eq:S-Matrix-cluster} are not the same. If one writes explicitly the expression for the double scattering, one also finds a different combination of $(2\omega_i)^{1/2}$ factors than in the two former cases. In order to sum the different terms in the FCA series with the same normalization a redefinition of amplitudes is convenient and this was done in \cite{Roca:2010tf}. One simply has to define the $\tilde{t}_i$ matrices,
\begin{align}
  \tilde{t}_i=t_i\left(\frac{\omega(c)\omega^\prime(c)}{\omega(a_i)\omega^\prime(a_i)}\right)^{1/2}\approx t_i\frac{m_{\rm c }}{m_{ai}},~~ i=1,2.\label{Eq:normalization}
\end{align}
Together with the expression for the $G_0$ function given as
\begin{align}
 G_0(q^0)=\frac{1}{2M_{\rm c}}\int_{\mathbb{R}^3}\frac{{\rm d}^3q}{(2\pi)^3}F_R(\vec{q}^{\,2})\frac{1}{\left(q^0\right)^2-\vec{q}^2-m^2_{a_3}+{\rm i}\epsilon}
 \label{Eq:G0},
\end{align}
where $F_R(\vec{q}^{\,2})$ is the form factor of the cluster wave function that we describe below. With the normalization of Eq. \eqref{Eq:normalization} we have

\begin{align}
 T=T_1+T_2=\frac{\tilde{t}_1+\tilde{t}_2+2\tilde{t}_1\tilde{t}_2G_0}{1-\tilde{t}_1\tilde{t}_2G_0^2}
 \label{Eq:solution-FCA}.
\end{align}
Since in this case $\tilde{t_1}=\tilde{t_2}$, the former expression can be simplified and we find:
\begin{align}
  T=\frac{2\tilde{t}_1}{1-\tilde{t}_1G_0}.
 \label{Eq:solution-FCA-simplified}
\end{align}

The $F_R$ function in Eq. (\ref{Eq:G0}) is the form factor of the resonance or cluster, which is the Fourier transform of its wave function. There $q^0$ and $m_{a_3}$ are the energy and the mass of the $\rho$ meson. We will consider the following form factor description for s-wave functions \cite{Roca:2010tf,Bayar:2015oea,YamagataSekihara:2010pj}, 
\begin{align}
 F_R(\vec{q}^{\,2})=\frac{1}{N}\int_{\Omega}d^3p \mathcal{A}(\vec{p})\mathcal{A}(\vec{p}-\vec{q}),~~\Omega\coloneqq\left\lbrace \left|\vec{p}\right|,\left|\vec{p}-\vec{q}\right|<\Lambda\right\rbrace
 \label{Eq:Form-factor-formula-1},
\end{align}
where $\mathcal{A}$ and $N$ are defined by,
\begin{align}
 \mathcal{A}(\vec{p})&=\frac{1}{M_{\rm c}-\omega_{a_1}(\vec{p})-\omega_{a_2}(\vec{p})}\label{Eq:Form-factor-formula-2},\\
 N &=F_R \left(\vec{q}^{\,2}=0\right),\label{Eq:Form-factor-formula-3}
\end{align}
and $\Lambda$ is a three-momentum cutoff used to regularize the $B^\ast\bar{B}^\ast$ $G$ function in the $B^\ast$ system in order to obtain the $B^\ast\bar{B}^\ast$ as a bound state \cite{YamagataSekihara:2010pj}.

Eqs. (\ref{Eq:solution-FCA}) and \eqref{Eq:solution-FCA-simplified} have as argument the total center of momentum (c.o.m.) invariant energy $\sqrt{s}$ and, in analogy with \cite{Bayar:2015oea}, we would like to share this energy with the two particle subsystems. We follow the procedure of Ref. \cite{Bayar:2015oea} and we write here the expressions of the energy for the $\rho \bar{B}^*$ and $\rho B^*$ two-body systems $\sqrt{s_{1(2)}}$ in terms of $\sqrt{s}$,
\begin{align}
 s_{1(2)}=(p_{a_3}+p_{a_{1(2)}})^2=(E_3+E_{1(2)})^2-\vec{P}_{2(1)}^{\,2}.
 \label{Eq:two-body-energy-12}
\end{align}
The total energy is shared between the particles proportionally to their masses,
\begin{align}
  E_{i}\left(\sqrt{s}\right)&=\sqrt{s} \frac{m_i}{m_{a_3}+M_{\rm c}},~~i=1,2,3,~~m_3= m_{a_3}\label{Eq:Energy-two-body-1},\\
  m_{1(2)}&=m_{a_1(a_2)}\frac{M_{\rm c}}{m_{a_1}+m_{a_2}} \label{Eq:Energy-two-body-2}  ,
\end{align}
where $m_{a_j}$ is the mass of the $a_j$ particle. The total three-momentum of the two particle system appearing in Eq. (\ref{Eq:two-body-energy-12}) is estimated in terms of the binding energy of $a_1$ and $a_2$ in the cluster, i.e., 
\begin{align}
  \vec{P}^{\, 2}_{1(2)} \approx& 2m_{a_{1(2)}}B_{1(2)} \label{Eq:three-momentum-two-body},\\
  B_{1(2)}=& E_{1(2)}\left(\sqrt{s}=M_{\rm c}+m_{a_3}\right)-E_{1(2)}\left(\sqrt{s}\right). \label{Eq:Binding-energy}
\end{align}
With Eqs. \eqref{Eq:two-body-energy-12}-\eqref{Eq:Binding-energy} we are ready to implement the two-body $T$-matrix elements $\hat{t}^{\,I}_j$ in Eq. \eqref{Eq:t_j-I-1} in order to solve Eq. \eqref{Eq:solution-FCA-simplified}.

\subsection{The two-body interaction \label{Sec:two-body}}
As we have mentioned in Sec. \ref{Sec:FCA-F-E}, the $\rho B^*$ \footnote{And henceforth the $\rho \bar{B}^*$, so $\hat{t}_1=\hat{t}_2$} interaction has been studied in \cite{Soler:2015hna} in the scheme of the local Hidden Gauge approach \cite{Bando:1984ej}, with unitarization of the amplitudes in coupled channels solving the on-shell version of the factorized Bethe-Salpeter equation,
\begin{align}
 \hat{t}^{\,I,J}=\left[\mathds{I}_{2\times 2}-V^{I,J}G\right]^{-1}V^{I,J}
 \label{Eq:B-S-EQ},
\end{align}
where $I$ is the two-body total isospin and $J$ the two-body spin. The $(V)_{i,j}$ matrix elements are the tree level amplitudes where we label $1,1$ as $\rho B^*\rightarrow \rho B^*$, $1,2$ as $\rho B^*\rightarrow \omega B^*$ and $2,2$ as $\omega B^*\rightarrow \omega B^*$ interaction. $\mathds{I}_{2\times 2}$ is the $2\times 2$ identity matrix. The $G$ is a diagonal matrix which contains the loop function for a $\rho B^*$ and an $\omega B^*$ loop,
\begin{align}
 (G)_{j,j} &={\rm i}\int_{\mathbb{R}^4}\frac{{\rm d}^4q}{(2\pi)^4}\frac{1}{q^2-\hat{m}_{j}+{\rm i}\epsilon}\frac{1}{(P-q)^2-m_{B^*}+{\rm i}\epsilon},~~j=1,2\notag\\
 \hat{m}_{1(2)} &=m_{\rho(\omega)}.
 \label{Eq:loop-matrix}
\end{align}
We have employed a three-momentum cutoff $q_{\rm max}$ to regularize the loops of \eqref{Eq:loop-matrix}.   The four vector $P^\mu$ is the total four-momentum of the two-particle system $s_{1(2)}=P^2$. The cutoff was fixed around $q_{\rm max}=1300$ MeV in order to reproduce an already known state, the $B^*_2(5747)$. By solving Eq. \eqref{Eq:B-S-EQ} we get the $T$-matrix element for the $\rho B^*\rightarrow \rho B^*$ interaction in $I=1/2,3/2$ and $J=2$. 

In the two-body interaction, the width of the states has been also investigated. In order to do this we have convoluted the loop function considering the $\rho$ meson width, and also we have included the imaginary part of the box diagrams. The parameters of the two-body interaction in the present work are set equal to those mentioned in \cite{Soler:2015hna}.

Further details of this section can be found in the works of Refs. \cite{Soler:2015hna,Molina:2009eb}. In the next section we will show the results that we obtain when we apply our formalism to the $\rho B^*\bar{B}^*$ system.

\section{Results}
\subsection{Bound state in the FCA to Faddeev equations}
In Sec. \ref{Sec:FCA-F-E} we have stated that we would use the form factor of Eqs. \eqref{Eq:Form-factor-formula-1} to \eqref{Eq:Form-factor-formula-3} to describe the clustering effect in the propagator of the $\rho$ meson \eqref{Eq:G0}. As we can see in Eq. \eqref{Eq:Form-factor-formula-1} the form factor is regularized by a three-momentum cutoff $\Lambda$ which, as mentioned before, is the same one used in the regularization of the $B^\ast \bar{B}^\ast$ loops in the study of the $B^\ast\bar{B}^\ast$ interaction in \cite{Ozpineci:2013qza}. In that work, values of $\Lambda$ around $415-830$ MeV were used. We take a range of values here for $500-1200$ MeV and see how the results change with $\Lambda$. 
\begin{figure}[h!]
 \includegraphics[scale=0.35]{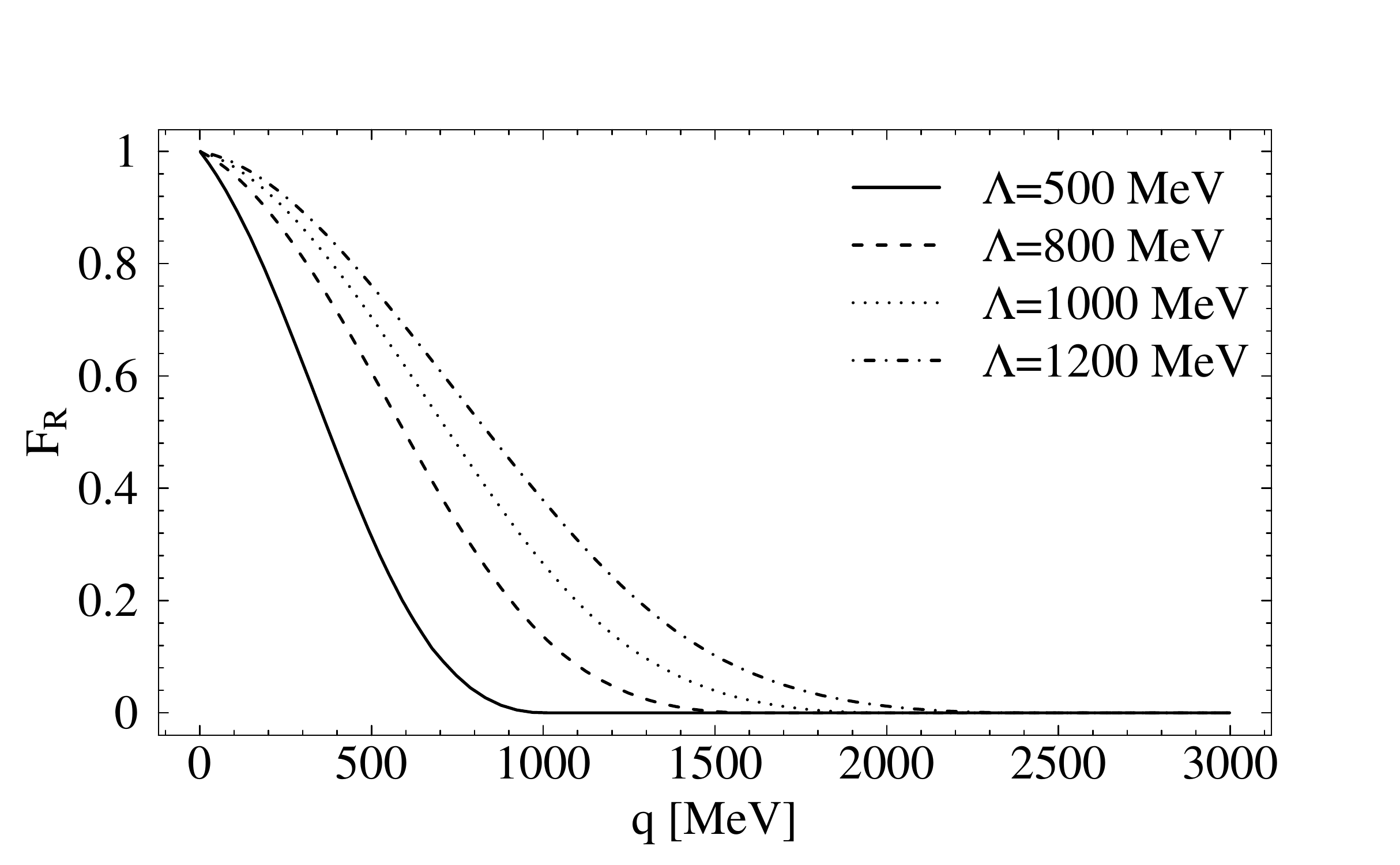}
 \caption{Form factor of Eq. \eqref{Eq:Form-factor-formula-1} in terms of $q= |\vec{q}|$ for different values of the three-momentum cutoff $\Lambda$.}
 \label{Fig:Form-factor}
\end{figure}
\begin{figure}[h!]
 \includegraphics[scale=0.35]{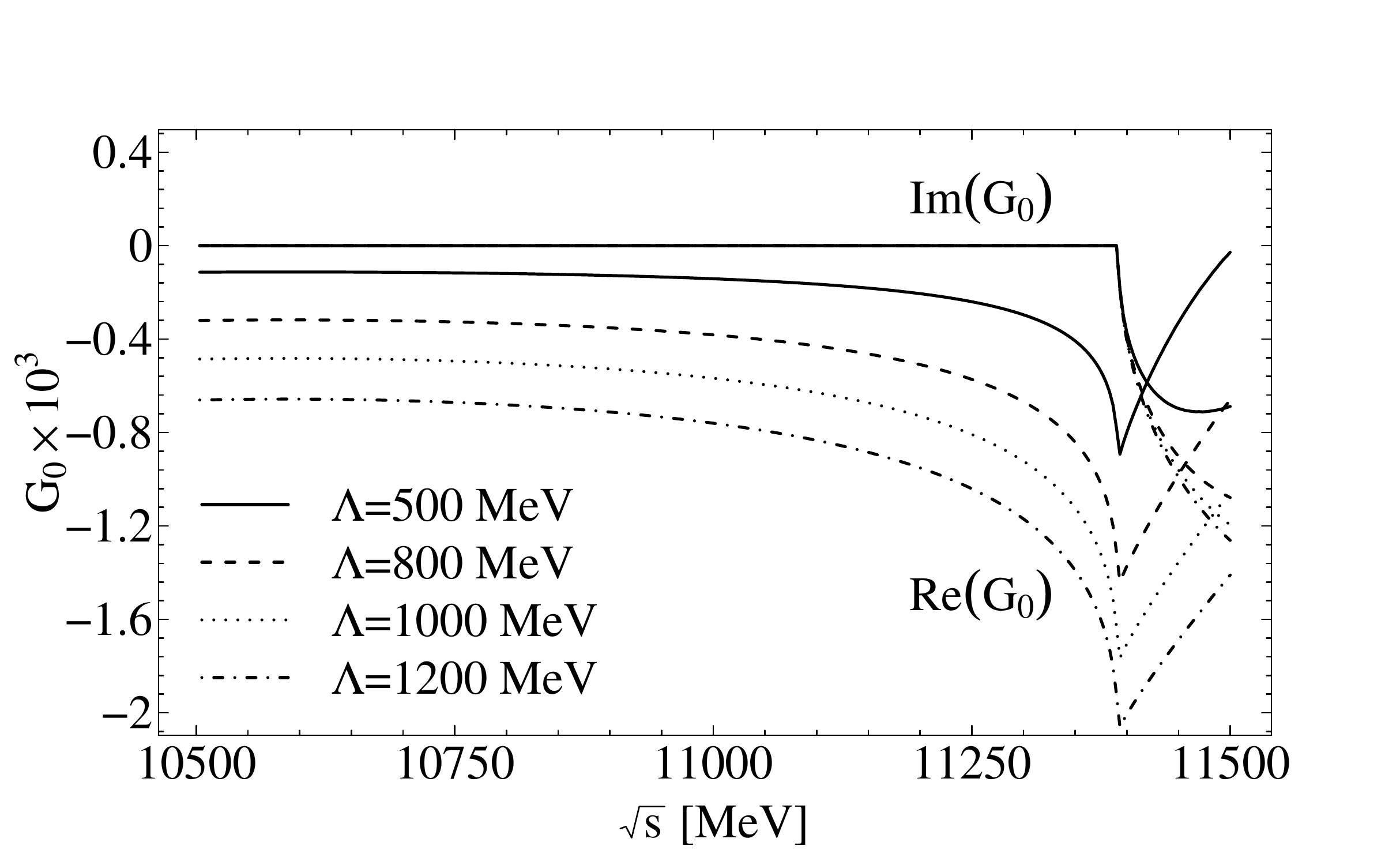}
 \caption{The energy dependence of the $\rho$ propagator function of Eq. \eqref{Eq:G0} for different values of the cutoff $\Lambda$.}
 \label{Fig:G0}
\end{figure}
\begin{figure}[h!]
 \includegraphics[scale=0.35]{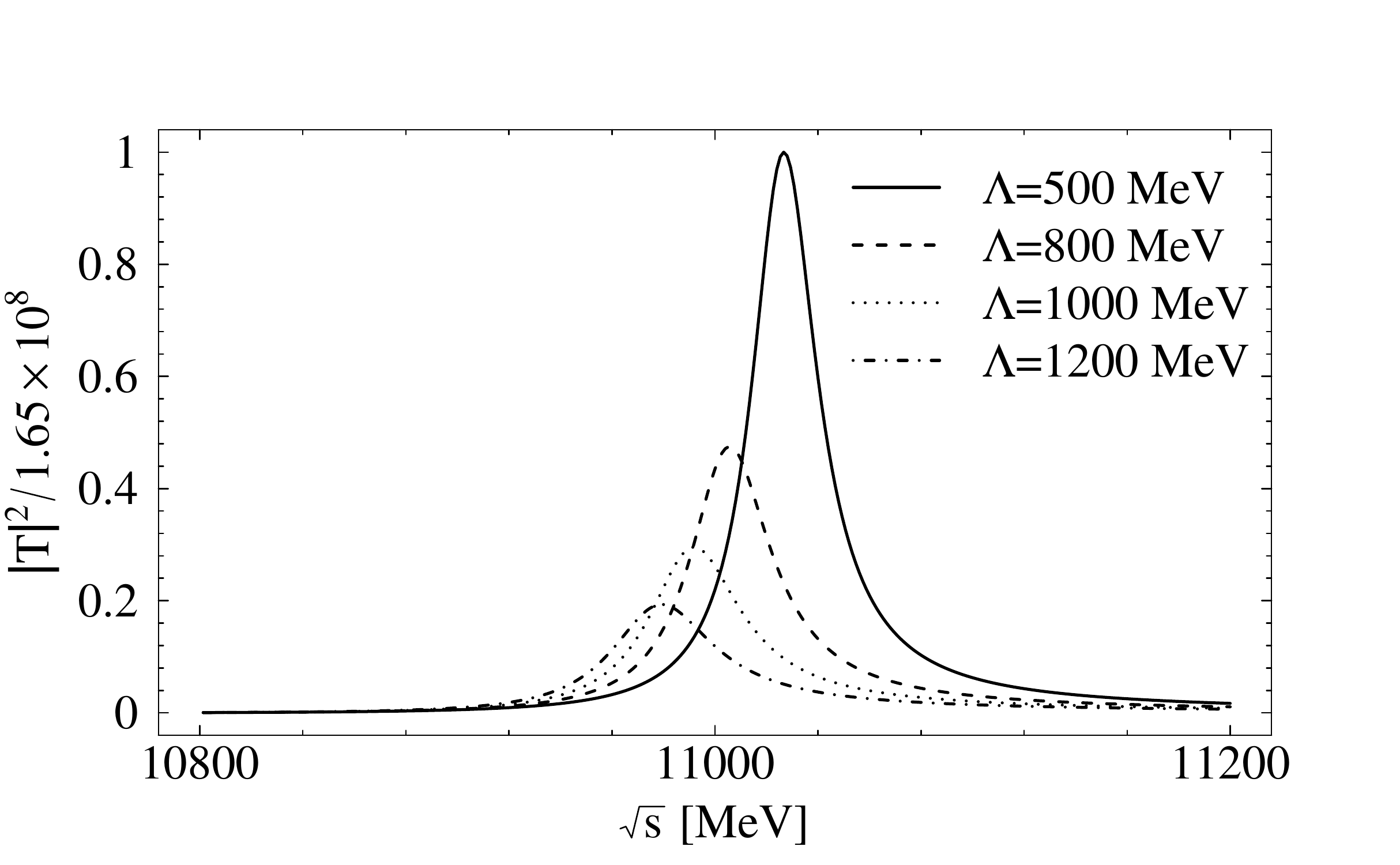}
 \caption{The energy dependence of the solution of \eqref{Eq:solution-FCA-simplified} for different values of the cutoff $\Lambda$.}
 \label{Fig:T2-1}
\end{figure}
In Fig. \ref{Fig:Form-factor} we can see the influence of the cutoff $\Lambda$ in the form factor. This dependence is translated into the propagator $G_0$ of Eq. \eqref{Eq:G0}. In Fig. \ref{Fig:G0} we plot the $G_0$ as function of the energy for different values of $\Lambda$. 

We consider the $\rho$-$(B^*\bar{B}^*)$ cluster system in order to fix the energy of the $\rho$ meson in terms of the total c.o.m. energy,
\begin{align}
 q^0=\frac{s+m_\rho^2-M^2_{\rm c}}{2\sqrt{s}}.
 \label{Eq:rho-meson-energy}
\end{align}
We recall here that we have chosen the value of the cluster mass as $M_{\rm c}=10616$ MeV, \cite{Ozpineci:2013qza}.
We have used these different values of the cutoff in order to solve Eq. \eqref{Eq:solution-FCA-simplified}. As we can see in Fig. \ref{Fig:T2-1}, we find a peak in the $|T|^2$ energy distribution, close to $11000$ MeV using different values of $\Lambda$. This result can be identified with a new $J=3$ state as a consequence of the three-body interaction. In Tab. \ref{Table:Mass-width-J3} we show the mass and width associated with this new state for different values of the $\Lambda$ parameter. 

A change of $200$ MeV in the cutoff brings up a change of the order of $10$ MeV in the mass, and of $4$ MeV in the width. Thus, these values are quite stable with respect to changes in the values of the $\Lambda$ parameter. The range of values of $\Lambda$ taken, in line with those used in \cite{Ozpineci:2013qza}, are of the order as those used in studies of related systems \cite{Bayar:2015oea} and can be considered of natural size \cite{Oller:2000fj}.
\begin{table}[t]
\centering
\caption{The mass and width of the $I(J^{PC})=1(3^{--})$ state found in the $\rho B^*\bar{B}^*$ interaction in the FCA to Faddeev equations.}
\label{Table:Mass-width-J3}
\begin{tabular}{ccccccccccccc}
\toprule[1pt]
$\Lambda$ [MeV]     & Mass [MeV]  & Width [MeV] \\
\midrule[0.5pt]
$500$	 &$ 11026$ & $30$\\
$800$	 &$ 11005$ & $38$\\
$1000$   &$10992$&$44$\\
$1200$   &$10978$&$48$ \\
\bottomrule[1pt]
\end{tabular}
\end{table}
The threshold of our three-body system is located at $M_{\rm c}+m_\rho=11391$ MeV. Thus, the bound state found has a binding energy of approximately $385$ MeV, taking $\Lambda=800$ MeV. The $\rho B^*$ interaction of spin 2 is also very attractive, in fact, the $B^*_2(5747)$ has a binding energy of $355$ MeV. The binding of the $B^*\bar{B}^*$ state is of $34$ MeV. We can conclude that the three-body interaction of the $\rho B^*\bar{B}^*$ system is more bound than either pair, as consequence of the combination of these two subsystems.

\subsection{Analysis of uncertainties}
In order to make a more detailed study of our model, we will show the results that we obtain when we modify some of the assumptions made in Sec. \ref{Sec:FCA-F-E}. First of all, the two-body energy has been estimated under the assumption of a particle energy distribution proportional to the masses (see Eqs. \eqref{Eq:Energy-two-body-1} and \eqref{Eq:Energy-two-body-2}). Let us denote it as scheme A. We will try another different way (scheme B) to share the total energy, as in Ref. \cite{YamagataSekihara:2010qk,Bayar:2015oea},
\begin{align}
 s_j=m_{a_3}^2+m_{a_j}^2+\frac{1}{2 M_{\rm c}^2}\left(s-m^2_{a_3}-M^2_{\rm c} \right)\left( M^2_{\rm c}+m^2_{a_j}-m^2_{a_{i\neq j} }\right).
 \label{Eq:Energy-two-body-3}
\end{align}
Additionally to Eq. \eqref{Eq:Energy-two-body-3}, we will also compute results using another approach. We will consider the $\rho$ $(B^*\bar{B}^*)$-cluster system as a two-body system so,
\begin{align}
 E_{3}=\frac{s+m_{a_3}^2-M_{\rm c}^2}{2\sqrt{s}},\label{Eq:Energy-two-body-4-1}\\
 E_{1,2}=\frac{1}{2}\frac{s+M_{\rm c}^2-m_{a_3}^2}{2\sqrt{s}} \label{Eq:Energy-two-body-4-2}.
\end{align}
Combining Eqs. \eqref{Eq:Energy-two-body-4-1} and \eqref{Eq:Energy-two-body-4-2} with Eqs. \eqref{Eq:two-body-energy-12}, \eqref{Eq:three-momentum-two-body} and \eqref{Eq:Binding-energy} we have another expression for the two-body energy, we will call it the scheme C. Let us consider the c.o.m. energy of the three-body system equals to $11000$ MeV for reference. Respectively, the different schemes A, B and C assign a $\rho B^\ast$ two-body energy of $5706$, $5727$ and $5692$ MeV, there is a difference of the order of 20 MeV. In Tab. \ref{Table:Mass-ABC} we can observe the differences in the peak position and the width caused by the different employed schemes.
\begin{table}[t]
\centering
\caption{The mass and width of the $I(J^{PC})=1(3^{--})$ state found in the $\rho B^*\bar{B}^*$ interaction considering three different schemes of energy distribution. $\Lambda$ is fixed to $900$ MeV.}
\label{Table:Mass-ABC}
\begin{tabular}{ccccccccccccc}
\toprule[1pt]
Scheme   & A & B & C \\
\midrule[0.5pt]
Mass [Mev]  &$10998$& $10977$ & $11012$\\
Width [MeV]   &$40$& $42$     & $40$\\
\bottomrule[1pt]
\end{tabular}
\end{table}

Another approximation used in the previous sections was to consider the normalization factor $\sqrt{2\omega(\rm{c})2\omega'(\rm{c})}/\sqrt{2 \omega(a_i)2\omega'(a_i)}$ as a constant. We have implemented this energy dependence in the three different cases of two-body energy distribution considered here, in order to have an idea of the influence in the results. The results are shown in Tab. \ref{Table:Mass-ABC-norm-factor}. 

As we can see in Tab. \ref{Table:Mass-ABC}, the change in the peak position depends on the way of sharing the three-body energy into the two-body systems. The difference between the three situations is a change of the order of 30 MeV in the peak position, while the width found is not much altered. The new results considering the energy dependent factor of Eq. \eqref{Eq:normalization}, Tab. \ref{Table:Mass-ABC-norm-factor}, induce another change of the same order in the mass for the schemes A and B. The width is also different, being of the order of ten MeV in comparison with the previous situation of Tab. \ref{Table:Mass-ABC}. The results suggest that the values found for this state are quite stable, since the mass obtained is roughly $(10949-11026)$ MeV and the width  $(30-53)$ MeV. The systematic uncertainties let a range of variation of less than one hundred MeV for the mass, which is an acceptable situation \cite{Bayar:2015oea}. Note that in all the cases a quasibound state always appears, so the prediction for the existence of this state is rather solid.

\begin{table}[t]
\centering
\caption{The mass and width of the $I(J^{PC})=1(3^{--})$ state found in the $\rho B^*\bar{B}^*$ interaction considering the energy dependent factor $ \sqrt{2\omega(\rm{c})2\omega'(\rm{c})}/\sqrt{2 \omega(a_{1(2)})2\omega'(a_{1(2)})}$ and three different schemes of energy distribution. $\Lambda$ is fixed to $900$ MeV.}
\label{Table:Mass-ABC-norm-factor}
\begin{tabular}{ccccccccccccc}
\toprule[1pt]
Scheme   & A & B & C \\
\midrule[0.5pt]
Mass [Mev]  &$11017$& $10949$ & $11012$\\
Width [MeV]   &$33$& $53$     & $40$\\
\bottomrule[1pt]
\end{tabular}
\end{table}

\section{Summary}
In this work we have studied the three-body system which consists of a $\rho$ meson and a $B^\ast \bar{B}^\ast$ cluster. We have solved the Faddeev Equations in the Fixed Center Approximation using the two-body interaction obtained in the unitary local Hidden Gauge approach. The $B^\ast \bar{B}^\ast$ cluster is assumed to be the $J=2$, $I=0$ state found in the previous work of Ref. \cite{Ozpineci:2013qza}. Taking the $\rho B^\ast$ interaction in $J=2$ from \cite{Soler:2015hna} we find a $I(J^{PC})=1(3^{--})$ state of mass $10987\pm 40$ MeV and width $40\pm 15$ MeV. Related to this mass is the two particle c.o.m. energy of the $\rho B^\ast$ subsystem that we find close to the mass of the $B^\ast_2(5747)$ resonant state appearing in the $J=2$ $\rho B^\ast$ interaction. We can conclude that the presence of this resonance plays an important role in the dynamics of the generation of the new three-body state. 

\section*{Acknowledgments}
This work is partly supported by the Spanish Ministerio de Economia y Competitividad and
European FEDER funds under the contract number FIS2011-28853-C02-01 and FIS2011-28853-
C02-02, and the Generalitat Valenciana in the program Prometeo II-2014/068. We acknowledge
the support of the European Community-Research Infrastructure Integrating Activity Study
of
Strongly Interacting Matter (acronym HadronPhysics3, Grant Agreement n. 283286) under the
Seventh Framework Programme of EU.

\bibliographystyle{plain}

%\end{multicols}
\end{document}